\documentstyle[12pt]{article}

\bigskip
\title{ Remarks on the Star-Triangle Relation in the Baxter-Bazhanov Model}
\author{Zhan-Ning Hu \thanks{\bf email address: huzn@itp.ac.cn}
\\ Institute of Theoretical Physics, Academia Sinica \\
  P. O. Box 2735, Beijing 100080, China \thanks{\bf mail address}
\and
\centerline{Bo-Yu Hou~~~~}\\
\centerline{Institute of Modern Physics, Northwest University, Xian 710069,
China}}
\date{August 21, 1994, AS-ITP-94-39}
\begin{document}
\maketitle
\bigskip

\begin{abstract}
In this letter we show that the restricted star-triangle relation introduced
by Bazhanov and Baxter can be obtained either from the star-triangle relation
of chiral Potts model or from the star-square relation which is proposed by
Kashaev $et ~al$ and give a response of the guess which is suggested by
Bazhanov and Baxter in Ref.  \cite{b2}.
\smallskip
\bigskip

\bigskip


\end{abstract}
\newpage

\section{Introduction}
Recently much progress has been made in the three-dimensional integrable
 lattice models. Bazhanov and Baxter generalized the trigonometric
 Zamolodchikov model with two states \cite{z1} to the case of the arbitrary
$N$ states \cite{b1,b2}. The star-star relation and the star-square relation
 of this model are discussed in detail \cite{b2,k12,h12}. Boose and Mangazeev
$et~al$ enlarge the integrable lattice model in three dimensions to the case
 where the weight functions are parameterized in terms of elliptic functions
\cite{m1,ms,bm}. Just as the Yang-Baxter equations or the star-triangle
 relations play a central role in the theory of two-dimensional integrable
models,  the tetrahedron relations replace the Yang-Baxter equations as the
 commutativity conditions \cite{jm} for the three-dimensional lattice models.
 And the restricted star-triangle relations of the cubic lattice model
introduced  by Bazhanov and Baxter have the following form:
\begin{equation} \label{e1}
\sum^{N-1}_{l=0} \frac{w(v_2 , a-l)}{w(v_1, -l) \gamma(b,l)}=\varphi_1(v_1,v_2)
\frac{w(v'_2,-b) w(v_2/(\omega v_1),a)}{w(v'_1, a-b)},
\end{equation}
\begin{equation} \label{e2}
\sum^{N-1}_{l=0} \frac{w(v_3, -l) \gamma(b,l)}{w(v_4,a-l)}=\varphi_2(v_3,v_4)
\frac{w(v'_3,a-b)}{w(v'_4,-b) w(v_4/v_3,a)},
\end{equation}
where $\varphi_1$ and $\varphi_2$ are scalar functions and
\begin{equation}
\frac{w(v,a)}{w(v,0)}=[\Delta(v)]^a \prod^a_{j=1} (1-\omega^jv)^{-1},~~~
v^N+\Delta^N=1,
\end{equation}
\begin{equation}
\omega=exp(2\pi i/N), ~~\omega^{1/2}=exp(i\pi /N), ~~\gamma(a,b)=\omega^{ab}.
\end{equation}
$v_i$ and $v'_i (i=1,2,3,4)$ satisfy
\begin{equation} \label{e3}
v'_1 = \frac{v_2\Delta (v_1)}{\omega v_1\Delta (v_2)},~~
\Delta (v'_1)=\frac{\Delta (v_2/(\omega v_1))}{\Delta (v_2)},~~
v'_2=\frac{\Delta (v_1)}{\Delta (v_2)},~~
\Delta(v'_2)=\frac{\omega v_1\Delta(v_2/(\omega v_1))}{\Delta(v_2)};
\end{equation}
\begin{equation} \label{e4}
v'_3=\frac{v_4\Delta (v_3)}{v_3\Delta(v_4)},~~
\Delta(v'_3)=\frac{\Delta(v_4/v_3)}{\Delta(v_4)},~~
v'_4=\frac{\Delta(v_3)}{\omega\Delta(v_4)},~~
\Delta(v'_4)=\frac{v_3\Delta(v_4/v_3)}{\Delta(v_4)}.
\end{equation}
Eq. (\ref{e1}) and Eq. (\ref{e2}) can be changed each other. Bazhanov and
Baxter point out that it is quite possible that Eq. (\ref{e1}) is a
particular  case of a more general relation and $\gamma$ is just a limiting
value of a more complex function. The purpose of this letter is to give a
response of it. In Sec. 2 the star-triangle relation of Baxter-Bazhanov model
is obtained either from the star-triangle relation of the chiral Potts model
or from the star-square relation introduced by Kashaev $et~al$. In Sec. 3, the
result is changed into the form of Eqs. (\ref{e1}) and (\ref{e2}). It should
 be note that the last relation in Eqs. (\ref{e3}) is different from the
original one. The detail will be given also in Sec. 3.

\section{The Star-Triangle Relation of Baxter-Bazhanov Model}

As is well-known, the star-triangle relation of the chiral Potts model can be
formulated as
\begin{equation} \label{e21}
\sum ^N_{l=1} \bar{w}^{CP}_{qr}(m-l) w^{CP}_{pr}(n-l)\bar{w}^{CP}_{pq}(l-k)
=R_{pqr}w^{CP}_{pq}(n-m)\bar{w}^{CP}_{pr}(m-k)w^{CP}_{qr}(n-k)
\end{equation}
where
\begin{equation} \label{e20}
\frac{w^{CP}_{pq}(n)}{w^{CP}_{pq}(0)}=\prod^n_{j=1}\frac{d_pb_q-a_pc_q\omega^j}{b_pd_q-c_pa_q\omega^j},~~~
\frac{\bar{w}^{CP}_{pq}(n)}{\bar{w}^{CP}_{pq}(0)}=\prod^n_{j=1}\frac{\omega
a_pd_q-d_pa_q\omega^j}{c_pb_q-b_pc_q\omega^j}
\end{equation}
and
\begin{equation}
a^N_p+k'b^N_p=kd^N_p,~~k'a^N_p+b^N_p=kc^N_p,~~k^2+k'^2=1.
\end{equation}
Let
\begin{equation}
w(x,y,z|l)=\prod^l_{j=1}\frac{y}{z-x\omega^j},~~~
x^N+y^N=z^N,
\end{equation}
\begin{equation}
w_{pq}(n)\equiv w(\omega^{-1}c_pb_q,d_pa_q,b_pc_q|n)
\end{equation}
and define the map $R$ as
\begin{equation}
R: ~~~ (a_p,b_p,c_p,d_p)\longrightarrow (b_p,\omega a_p,d_p,c_p).
\end{equation}
When we set $a_p=d_r=0$, the following relations are obtained:
\begin{equation}
\frac{ w^{CP}_{pr}(n)}{ w^{CP}_{pr}(0)} = w_{pR(r)}(n),~~~~~~~
\frac{\bar{w}^{CP}_{pr}(n)}{\bar{w}^{CP}_{pr}(0)}=\frac{1}{w_{pr}(-n)};
\end{equation}
\begin{equation}
\frac{w^{CP}_{pq}(n)}{w^{CP}_{pq}(0)}=w_{pR(q)}(n),~~~~~~~
\frac{\bar{w}^{CP}_{pq}(n)}{\bar{w}^{CP}_{pq}(0)}=\frac{1}{w_{pq}(-n)};
\end{equation}
\begin{equation}
\frac{w^{CP}_{qr}(n)}{w^{CP}_{qr}(0)}=w_{R^{-1}(q)r}(-n),~~~
\frac{\bar{w}^{CP}_{qr}(n)}{\bar{w}^{CP}_{qr}(0)}=\frac{1}{w_{qr}(-n)}.
\end{equation}
By taking account of the star-triangle equation (\ref{e21}) of chiral Potts
model we get
\begin{equation} \label{e22}
\sum^N_{l=1}\frac{w_{pR(r)}(n-l)}{w_{qr}(l-m)w_{pq}(k-l)}=
R'_{pqr}\frac{w_{pR(q)}(n-m)w_{R^{-1}(q)r}(k-n)}{w_{pr}(k-m)}
\end{equation}
with $a_p=d_r=0$ where $R'_{pqr}$ is a scalar function. This is just the
star-triangle equation of Baxter-Bazhanov model. If we set $a_p=c_r=0$,
similarly we have
\begin{equation} \label{e23}
\sum^N_{l=1}\frac{w_{pR(r)}(n+l)}{w_{R(q)R(r)}(m+l)w_{pq}(k+l)}=
\bar{R'}_{pqr}\frac{w_{pR(q)}(n-m)w_{qR(r)}(n-k)}{w_{R(p)R(r)}(m-k)}
\end{equation}
where $\bar{R'}_{pqr}$ is also a scalar function. Both of the above two
equations can be changed into the form of Eqs. (\ref{e1}) and (\ref{e2}). It
will be discussed in Sec. 3. Now we give the connection between Eq.
(\ref{e22}) and the star-square relation in Baxter-Bazhanov model. Let
\begin{equation}
w(x,y,z|l)=(y/z)^lw(x/z|l),~~~\Phi (a-b)=\omega^{(a-b)(N+a-b)/2}.
\end{equation}
As the version of Kashaev $et~al$, the star-square relation can be written as
\cite{k12}
$$
\Bigg\{\sum_{\sigma\in
Z_N}\frac{w(x_1,y_1,z_1|a+\sigma)w(x_2,y_2,z_2|b+\sigma)}{w(x_3,y_3,z_3|c+\sigma)w(x_4,y_4,z_4|d+\sigma)}\Bigg\}_0 ~~~~~~~~~~~~~~~~~~~~~~~~~~~~~~~~~~~~~~~~~~~
$$
$$
=\frac{(x_2y_1/x_1z_2)^a(x_1y_2/x_2z_1)^b(z_3/y_3)^c(z_4/y_4)^d}{\Phi(a-b)\omega^{(a+b)/2}} ~~~~~~~~~~~~~~~~~~~~~~~~~~~~~~~
$$
\begin{equation} \label{e24}
{}~\times \frac{w(\omega
x_3x_4z_1z_2/x_1x_2z_3z_4|c+d-a-b)}{w\bigg(\frac{\displaystyle
x_4z_1}{\displaystyle x_1z_4}|d-a\bigg)w\bigg(\frac{\displaystyle
x_3z_2}{\displaystyle x_2z_3}|c-b\bigg)w\bigg(\frac{\displaystyle
x_3z_1}{\displaystyle x_1z_3}|c-a\bigg)w\bigg(\frac{\displaystyle
x_4z_2}{\displaystyle x_2z_4}|d-b\bigg)},
\end{equation}
where the subscript "0" after the curly brackets indicates that the l. h. s. of
the above equation is normalized to unity at zero exterior spins and the
constraint condition $y_1y_2z_3z_4/(z_1z_2y_3y_4)=\omega $ should be imposed
owing to spin $\sigma \in Z_N$ but r. h. s. of the above equation is
independent of $\sigma$. Set
\begin{equation}
\begin{array}{lll}
x_1=c_qb_r, & y_1=\omega d_qa_r, & z_1=b_qc_r,\\
x_2=c_pa_q, & y_2=d_pb_q, & z_2=b_pd_q,\\
x_3=\omega_{-1}c_pb_r, & y_3=d_pa_r, & z_3=b_pc_r,\\
x_4=0, & y_4=z_4. &
\end{array}
\end{equation}
By considering the "inversion" relation \cite{k12,h12}
\begin{equation}
\sum_{k\in Z_N}\frac{w(x,y,z|k,l)}{w(x,y,\omega
z|k,m)}=N\delta_{l,m}\frac{1-z/x}{1-z^N/x^N}
\end{equation}
where $\delta_{l,m}$ is the Kronecker symbol on $Z_N$ we get the equation
(\ref{e22}) from the star-square relation (\ref{e24}). Eq. (17) can be obtained
similarly.

\section{Discussion}
Firstly, Eq. (16) can be changed into the form of Eq. (\ref{e1}) and Eq.
(\ref{e2}) by using the notations
$$
v_1=\frac{\displaystyle c_pb_q}{\displaystyle \omega
b_pc_q},~~~v_2=\frac{\displaystyle b_qc_r}{\displaystyle
c_qb_r},~~~\Delta(v_1)=\frac{\displaystyle d_pa_q}{\displaystyle
b_pc_q},~~~\Delta(v_2)=\frac{\displaystyle \omega^{1/2}d_qa_r}{\displaystyle
c_qb_r},
$$
\begin{equation}
v'_1=\frac{\displaystyle a_qc_r}{\displaystyle
d_qb_r},~~~v'_2=\frac{\displaystyle c_pa_q}{\displaystyle
b_pd_q},~~~\Delta(v'_1)=\frac{\displaystyle \omega^{1/2}c_qa_r}{\displaystyle
d_qb_r},~~~\Delta(v'_2)=\frac{\displaystyle d_pb_q}{\displaystyle b_pd_q},
\end{equation}
$$
\Delta(\frac{\displaystyle v_2}{\displaystyle \omega v_1})=\frac{\displaystyle
\omega^{1/2}d_pa_r}{\displaystyle c_pb_r},
$$
and
\begin{equation}
\begin{array}{llll}
v_3=\frac{\displaystyle b_qc_r}{\displaystyle c_qb_r}, & v_4=
\frac{\displaystyle c_pb_q}{\displaystyle \omega b_pc_q}, & \Delta(v_3)=
\frac{\displaystyle \omega^{1/2}d_qa_r}{\displaystyle c_qb_r}, & \Delta(v_4)=
\frac{\displaystyle d_pa_q}{\displaystyle b_pc_q},\\
v'_3=\frac{\displaystyle d_qb_r}{\displaystyle \omega a_qc_r}, & v'_4=
\frac{\displaystyle b_pd_q}{\displaystyle \omega c_pa_q}, & \Delta(v'_3)=
\frac{\displaystyle c_qa_r}{\displaystyle a_qc_r}, & \Delta(v'_4)=
\frac{\displaystyle d_pb_q}{\displaystyle \omega^{1/2}c_pa_q};
\end{array}
\end{equation}
respectively, with $d_pb_r=\omega^{1/2}c_pa_r$.  $v_i$ and $v'_i~(i=1,2,3,4)$
 satisfy Eqs. (5) and (6). Here we show that the last relation in Eq. (5) is
 correct and this relation is different from the one  in Ref. \cite{b2}. Eq.
 (17) can  be also changed into Eq. (1) by setting
\begin{equation}
\begin{array}{llll}
v_1=\frac{\displaystyle b_pd_q}{\displaystyle \omega c_pa_q}, & v_2=
\frac{\displaystyle d_qa_r}{\displaystyle a_qd_r}, & \Delta(v_1)=
\frac{\displaystyle d_pb_q}{\displaystyle \omega^{1/2}c_pa_q}, & \Delta(v_2)=
\frac{\displaystyle c_qb_r}{\displaystyle a_qd_r}, \nonumber\\
v'_1=\frac{\displaystyle c_pb_q}{\displaystyle \omega b_pc_q}, & v'_2=
\frac{\displaystyle b_qd_r}{\displaystyle \omega c_qa_r}, & \Delta(v'_1)=
\frac{\displaystyle d_pa_q}{\displaystyle b_pc_q}, & \Delta(v'_2)=\
frac{\displaystyle d_qb_r}{\displaystyle \omega^{1/2}c_qa_r},
\end{array}
\end{equation}
$$
\Delta(\frac{v_2}{\omega v_1})=\frac{d_pb_r}{b_pd_r},
$$
with $c_pb_r=\omega^{1/2}d_pa_r$. Similarly Eq. (2) can be obtained easily
 from Eq. (17). In fact, each of the relations (16), (17) is a corollary of
 another by taking account of the "inversion" relation (21).

As a conclusion, in this letter we get the star-triangle relation of the
 Baxter-Bazhanov model from the star-triangle relation of the chiral Potts
 model and give a response to the guess proposed by Bazhanov and  Baxter. And
 the connection is found between the star-triangle relation and the
 star-square relation in the Baxter-Bazhanov model.

\section*{\bf Acknowledgment}

One of the authors (Hu) would like to thank H. Y. Guo, P. Wang and K.Wu for
the interesting discussions.


\newpage

\begin{thebibliography}{9}

\bibitem{z1} A. B. Zamolodchikov, {\it Commun. Math. Phys.} {\bf 79}: 489
(1981).

\bibitem{b1} V. V. Bazhanov and R. J. Baxter, {\it J. Stat. Phys.} {\bf 69}:
453 (1992).

\bibitem{b2} V. V. Bazhanov and R. J. Baxter, {\it J. Stat. Phys.} {\bf 71}:
839 (1993).

\bibitem{k12} R. M. Kashaev, V. V. Mangazeev and Yu. G. Stroganov, {\it Int. J.
Mod. Phys.} {\bf A8}: 587; 1399 (1993).

\bibitem{h12} Z. N. Hu, Three-dimensional star-star relation, {\it Int. J. Mod.
Phys.} (to appear); {\it Mod. Phys. Lett.} {\bf B8}: 779 (1994).

\bibitem{m1} V. V. Mangazeev, Yu. G. Stroganov, preprint IHEP 93-80,
(hep-th/9305145), {\it Mod. Phys. Lett.} {\bf A}, (to appear)

\bibitem{ms} V. V. Mangazeev, S. M. Sergeev and Yu. G. Stroganov, New series of
3D lattice integrable models, preprint, October (1993).


\bibitem{bm} H. E. Boos, V. V. Mangazeev and S. M. Sergeev, Modified
tetrahedron equations and related 3D integrable models, preprint, June (1994).

\bibitem{jm} M. T. Jaekel and J. M. Maillard, {\it J. Phys.} {\bf A15}: 1309
(1982).


\end{thebibliography}
\end{document}